\documentclass{cernyrep}
\usepackage{varwidth}
\usepackage{xcolor}

\begin{document}
\title{Medical Physics Commissioning}
\author{D. Meer}
\institute{Paul Scherrer Institut, Villigen, Switzerland}
\maketitle

\begin{abstract}
Medical commissioning is an important step in bringing a particle gantry into clinical operation for tumour treatments. This involves the parametrization and characterization of all relevant systems including beam delivery, patient table, imaging systems and connection to all required software components. This article is limited to the necessary tasks for the beam delivery system of a pencil beam scanning system. Usually, commissioning starts with the characterization of the unscanned beam and calibration of the beam energy. The next steps are the parametrization of the scanning system, the commissioning of the beam position monitoring system and characterization of the spot size, which all require precision better than \Unit{1}~{mm}. The commissioning effort for these tasks also depends on the gantry topology. Finally, calibration of the dose measurement system ensures that any dose distribution can be delivered with an absolute precision better than 1\%.
\end{abstract}


\section{Introduction}
Proton therapy systems are complex and commissioning is an important step in bringing such a facility into clinical operation. Medical physics commissioning is the intermediate step between technical commissioning and medical acceptance. Technical commissioning includes the installation and functional testing of system components, \eg the beam line magnets or the rotating mechanical gantry structure.

Within \emph{medical physics commissioning}, the main focus is on integral system tests and parametrization of the machine characteristics. The correct interaction between different system components is verified and important machine parameters are measured. This task not only consists of checking the beam scanning system and monitoring system, but also includes checking the mechanical systems (gantry rotation), the patient positioning systems, the different imaging systems and the transformation of coordinate system between them, as well as the software connection to the treatment planning software (TPS) and oncology information system. Finally, a properly configured safety system ensures a safe operation. After this important step, the system must be able to precisely deliver a predefined dose distribution with a given dose to a target.

What follows is medical acceptance of the whole treatment unit when operated by a responsible end-user. The system specifications are validated by numerous tests, performed by medical physicists. Many of these tests have end-to-end character and successful completion is a prerequisite before clinical operation can be started.

There is no strict separation between technical and medical commissioning, and responsibilities between producer and end-user might shift from one installation to another. It also makes a difference whether an established commercial product is installed in a clinical environment or a system is developed for clinical application at a research institute. In this report we look at the latter case, mainly reporting on the experience gained in the commission of the second gantry at the Paul Scherrer Institut (PSI). Gantry~2~\cite{Pedroni2004a} is an active scanning system and we therefore only concentrate on pencil beam scanning (PBS) rather than passive scattering. We mainly focus on the dose delivery system, although medical commissioning comprises much more than this, as mentioned before. In the first section, we discuss the commissioning of the unscanned beam. The following section explains the lateral beam spreading with the scanning system and the last section describes the commissioning tasks for the dose-monitoring system.

\section{Commissioning the unscanned beam}
Almost all particle therapy systems use beam transfer lines to transport the particles from the source (the accelerator) to the treatment rooms. Dipole and quadrupole magnets along the beam line help to direct and focus the particle beam to the isocentre. Additionally, a collimator, mechanical slits or scattering foils help to create the required phase space at the treatment location. Today, the vast majority of PBS systems do not use local energy modulation right in front of the patient but rely on variable beam energy from the accelerators or on upstream energy degrading systems.

\subsection{Beam line tuning}
To cover the full clinical range in depth, at least 50 different beam energies must be transported from the accelerator to the isocentre, and a beam line usually contains more than 30 devices. To optimize the beam line settings, well proven tools like TRANSPORT\cite{Brown1973} or Mad-X\cite{Herr2004} are available.

From a particular solution for the magnet settings for a given energy, beam line settings for other energies can be obtained by scaling the magnet currents with the particle momentum. This approach also gives smooth behaviour of the beam characteristics over the entire energy range. However, a linear model, as used in TRANSPORT, cannot describe the entire beam line, which also contains vacuum windows, collimators and drift distances in air or energy-absorbing materials. In order to obtain a full beam line description and a qualitative number for the beam line transmission, these models must be combined with Mote Carlo methods. Several beam profile monitors along the beam line can help to validate the predictions from such models.

\begin{figure}[ht]
\begin{center}
\includegraphics[width=.6\linewidth]{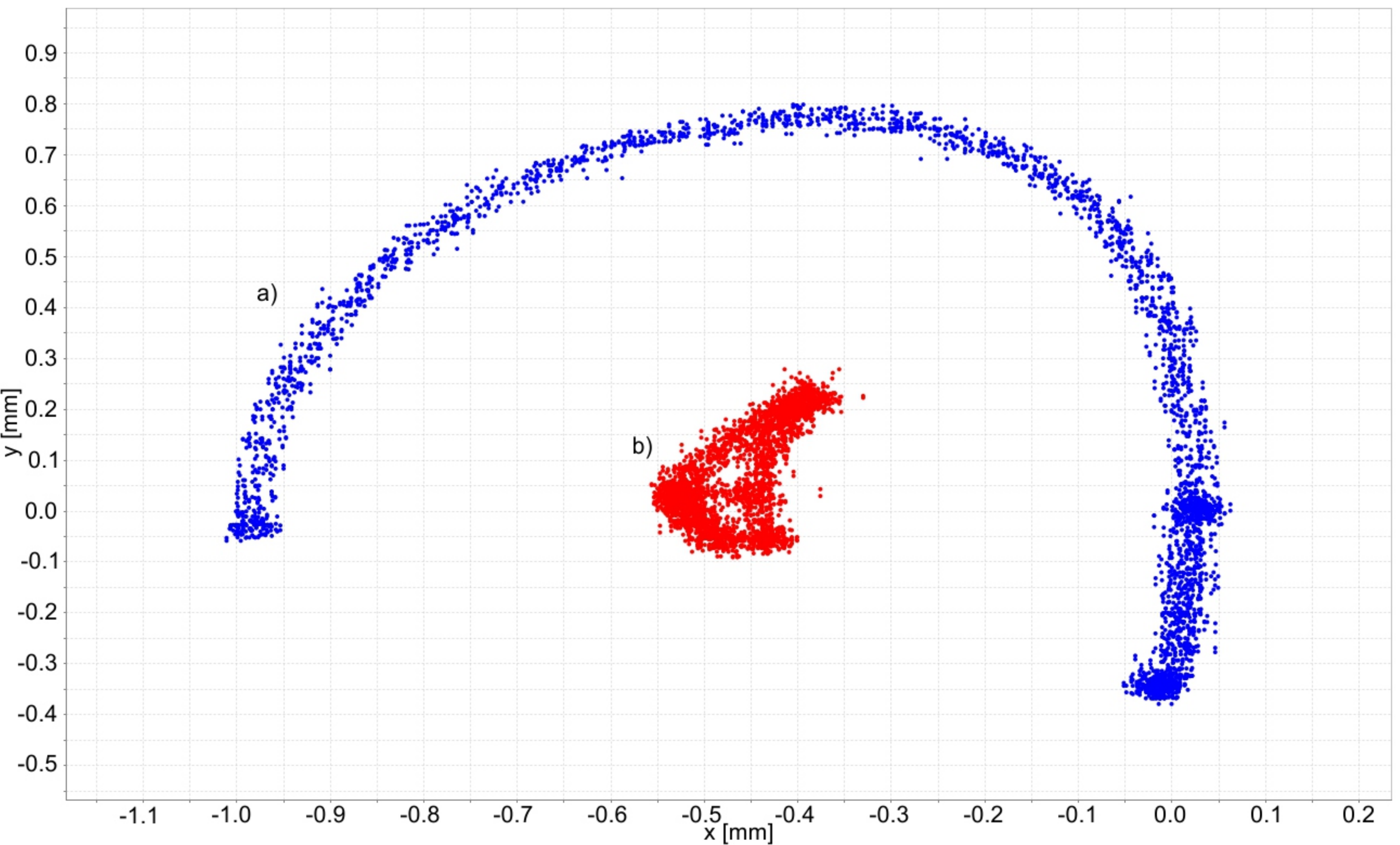}
\caption{Beam centring at the beginning on the rotating part of the gantry. The beam position is recorded continuously while the gantry rotates. The initial beam positions (a) and beam positions after proper centring (b) are shown. Maximum deviation is about \Unit{350}{$\mu$m} after optimization. }
\label{fig:BeamCentering.pdf}
\end{center}
\end{figure}

To complete the beam tuning, some manual optimization with steering magnets is necessary in order to centre the beam. It is of particular importance that the beam is well centred on the mechanical rotation axis of the gantry. This procedure can be simplified if a position monitor on the still horizontal but rotating part of the gantry is available and data from this monitor can be read while the beam is turned on and the gantry is rotating. If a circle is visible on this position monitor, the beam centring can still be improved as shown in \Fref{fig:BeamCentering.pdf}.
Steering magnets on the gantry also help to centre the beam. If necessary, gantry angle dependent solutions must be found.

\subsection{Energy calibration}
The range of the particles in matter (water) is one of the crucial parameters for TPS and needs to be adjusted with an absolute precision better than \Unit{\pm0.5}{mm}. Nevertheless, the calculations of beam tuning tools are based on beam momentum and, due to several factors, the comparison of the converted particle range with the measured range at the beam line often shows not negligible differences. Therefore, energy-to-range calibration is necessary and is performed with help of Bragg peak measurements. The range of a particle beam is defined as the 80\% point of the maximum ($R_{80}$) in the distal fall-off of the Bragg peak. By this definition, the range is insensitive to different momentum spread~\cite{Hsi2009}.

The preferred tools for measuring range in water are so-called range scanners. They consist of two dose-measurement chambers (plain parallel ionization chambers (ICs)): one is fixed at the entrance to the measurement phantom and the second one moves at depth. The ratio between the two chambers is calculated  at different depths to obtain the depth--dose curve. Commercial measurement tools such as the PEAKFINDER Water Column from PTW-Freiburg, Germany, are available for such measurements. The typically requested precision for range measurements is better than \Unit{0.2}{mm}, which corresponds approximately to a resolution of \Unit{0.1}{MeV}. At PSI, we have developed a water scanner which only consists of the movable chamber. We have access to the last dose monitor on the gantry which is used for data normalization.

\begin{figure}
\begin{center}
\includegraphics[width=.5\linewidth]{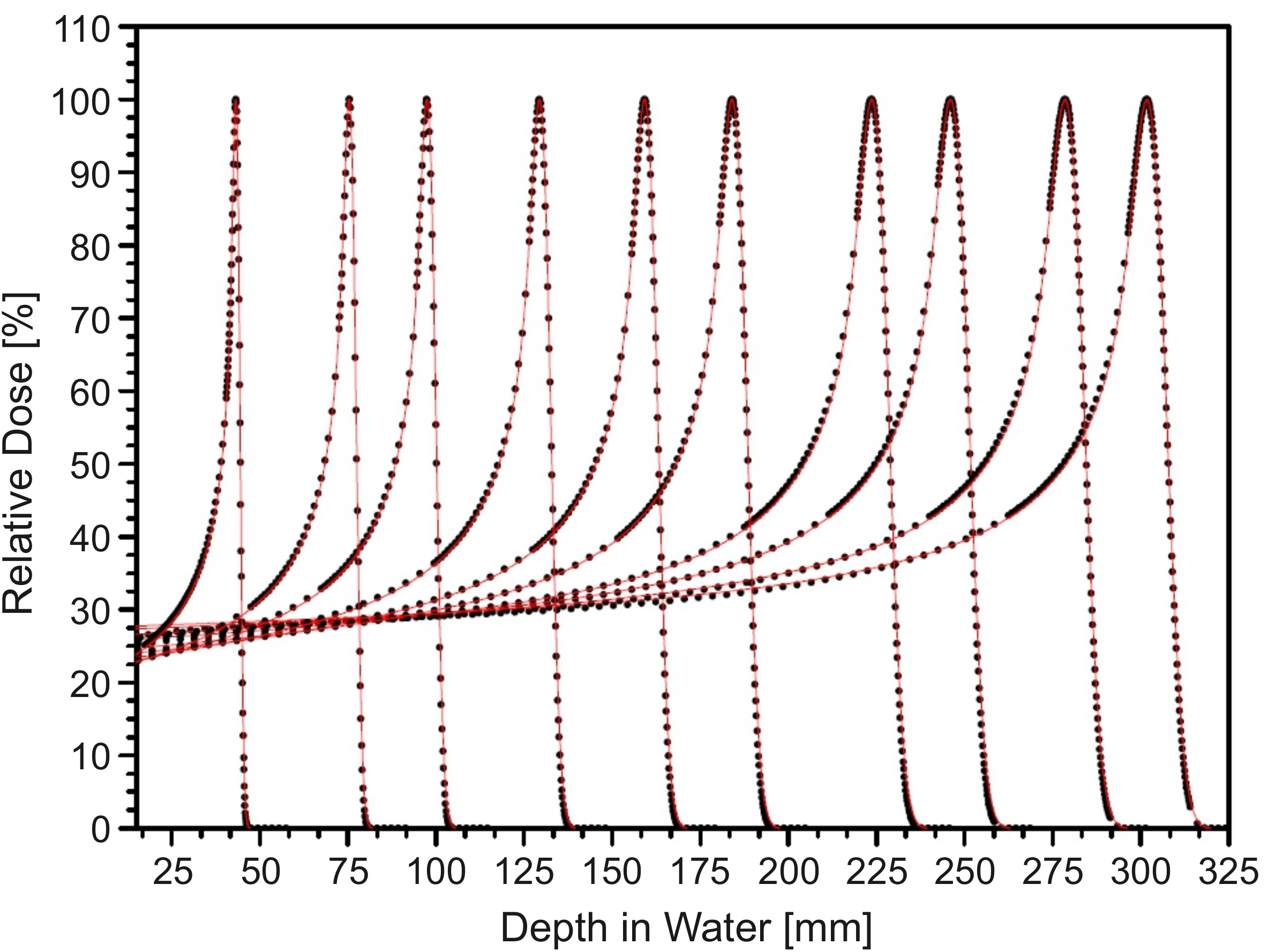}
\caption{Measured integral depth--dose curves for 10 different beam energies. The adaptive measurement granularity provides high resolution and efficient data acquisition.}
\label{fig:DepthDoseCurve.pdf}
\end{center}
\end{figure}

For determination of the $R_{80}$, it is sufficient to only measure the region around the Bragg peak with high resolution. However, the integral depth--dose profile is an important input for the TPS and must also be measured with high precision. As shown in \Fref{fig:DepthDoseCurve.pdf}, an adaptive granularity can help to accelerate measurements.

\subsection{Beam line supervision}
In systems with variable beam energy, the beam line settings are changed with high frequency and online supervision helps to verify the correct settings. Of particular importance are the dipoles of the energy-selection system, collimators and slits. The correct settings for the dipoles can be verified by an independent measurement of the magnetic field with Hall probes. The signal of the Hall probe is proportional to the $B$ field
\begin{equation}
V_\mathrm{Hall Probe} \propto B\propto p c= \sqrt{E_{\textrm{kin}}^ 2+2 E_{\textrm{kin}}E_0}~,
\end{equation}
and can be parametrized as a function of beam energy $E$,
\begin{equation}
V_\mathrm{Hall Probe} = a_1\sqrt{(E-a_2)^2+2E_0(E-a_2)}+a_3
\end{equation}
with three parameters $a_1$, $a_2$ and $a_3$. At PSI, the stability of the electronics limits the energy resolution of the beam line supervision to about \Unit{1}{MeV}, corresponding to approximately \Unit{2}{mm} range uncertainty in water.

\section{Calibration of the scanning system}
Once the unscanned beam has been tuned, commissioning of the scanning system can be started. First, the impact of the gantry topology on the calibration of the scanner magnets is discussed.

\subsection{Up-stream and down-stream scanning}
\begin{figure}[ht]
\begin{center}
\includegraphics[width=.8\linewidth]{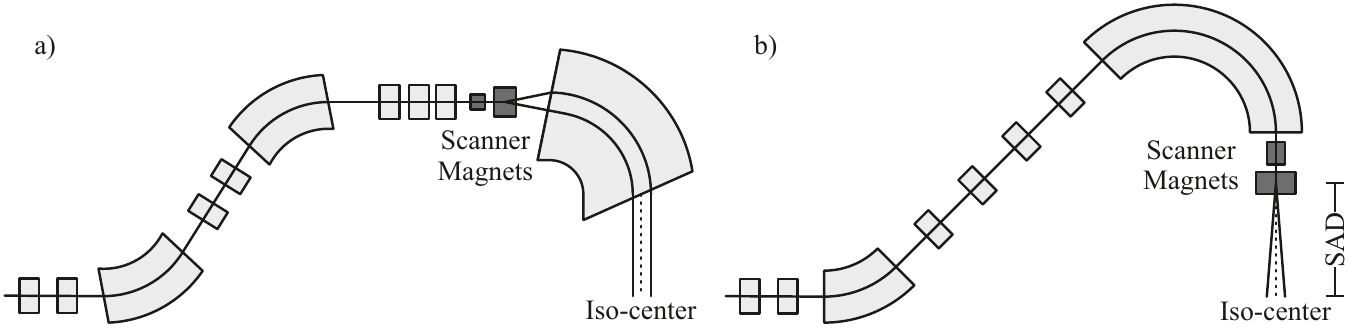}
\caption{Two possible gantry topologies with up-stream (a) and down-stream (b) scanning}
\label{fig:UpStreamScanning.pdf}
\end{center}
\end{figure}

The location of the scanner magnets on a PBS gantry has significant implications for both gantry design and commissioning. In a conventional design, the scanner magnets are the last devices in the beam line as shown in \Fref{fig:UpStreamScanning.pdf}(b). In this so-called \emph{down-stream scanning}, the beam direction is divergent while the beam is scanned laterally. The scanner magnets are placed at least \Unit{2}{m} away from the isocentre since a small source-to-axis distance (SAD) is unfavourable, leading to an increased skin dose, for example. To first order  there is a linear correlation between the spot position at the isocentre and the scanner magnet current. Another advantage is that the spot shape is unaffected for different scan positions. On the other hand, the calibration of the scanning system is sensitive to the longitudinal alignment of the device for the position measurement in the case of a finite SAD.

\emph{Up-stream scanning} is the other gantry topology in which the scanner magnets are installed before the last bending dipole as shown in \Fref{fig:UpStreamScanning.pdf}(a). In this configuration, a parallel beam (infinite SAD) can be obtained by properly optimizing the beam optics and the last bending dipole. A parallel scanned beam at the isocentre simplifies treatment planning and measurements for the quality assurance program. However, there are also additional technical challenges: beam-energy dependent inhomogeneities and fringe fields of the dipole can affect spot shape at the isocentre and beam focus needs to be dynamically adjusted when scanning the beam in the divergent plane. Consequently, higher-order corrections for position-to-current conversion need to be considered as well. PSI Gantry~2~\cite{Pedroni2011} is one of the few upstream scanning systems, and the calibration of that is discussed in the following section.

\subsection{Scanner magnet calibration}
The important prerequisite for accurate calibration of the scanner magnets is the availability of an appropriate device to measure the beam position at the isocentre for the full scanning range and different gantry angles. We decided to place a copy of the nozzle strip chamber on a rotatable support at the isocentre (see \Fref{fig:NozzleChamberCalibration.pdf} (right)). As a consequence of a parallel beam, the scan range at the isocentre is the same as in the nozzle, and we could therefore use a spare position monitor from the nozzle. This has the additional advantage that the standard read-out from the therapy control system (TCS) can be used, and data from the isocentre measurement can be integrated easily into the TCS logging file. The rotational support can turned remotely and integrates a beam dump which greatly simplifies the measurement at different gantry angles.

An alternative setup is to use a measurement device that can be attached to the gantry nozzle. In that case, the deformation of the supporting structure for different gantry angles needs to be considered.

The result of the sweeper calibration is two functions which provide the scanner magnet current
\begin{equation}
I_{\textrm{S}}^x = f^x(E,x,y)\ ,\ I_{\textrm{S}}^y = f^y(E,x,y)
\end{equation}
as a function of beam energy and lateral position at the isocentre for both scanner magnets. These functions can be look-up tables with additional interpolation or an analytic function. We decided to use polynomial functions that would be less sensitive to measurement errors and to give a smooth position dependency. The functions are calculated for every \Unit{10}{MeV} and interpolation is used for intermediate energies. Unfortunately, it was not possible to find one single polynomial expression for the full scanning range because each corner of the scan range showed substantial and specific deviation from the linear position to current correlation. To better consider the peculiarity of each corner, the global polynomial model was expanded with a local solution for each quadrant:
\begin{equation}
\begin{split}
I_{\textrm{S}}^x(x,y) & = \overbrace{g_1 x + g_2 y + g_3 x^2 + g_4 y^2 + g_5 y^3 + g_6 y^4}^{\text{Global function}} + \\
& \sum_{q=1}^{4} \underbrace{l_1^q xy + l_2^q x^2y + l_3^q xy^2 + l_4^q x^3y + l_5^q x^2y^2 + l_6^q xy^3 + l_7^q x^4y + l_8^q x^3y^2 + l_9^q x^2y^3 + l_{10}^q xy^4}_{\text{Local function for each quadrant}}
\end{split}
\end{equation}
and an analogue expression for $I_{\textrm{S}}^y$.

The ansatz with only mixed terms in $x$ and $y$ for the local solution ensures that the local contribution is vanishing when $x\rightarrow 0$ or $y\rightarrow 0$, and this provides a smooth function of position. The position to current calibration with this model requires 46 parameters for one beam energy. This is still much less data than a look-up table to deliver beam spots with a precision in the order of \Unit{100}{$\mu$m}.

\subsection{Position projection to the isocentre}
\begin{figure}[ht]
\begin{center}
\includegraphics[width=.3\linewidth]{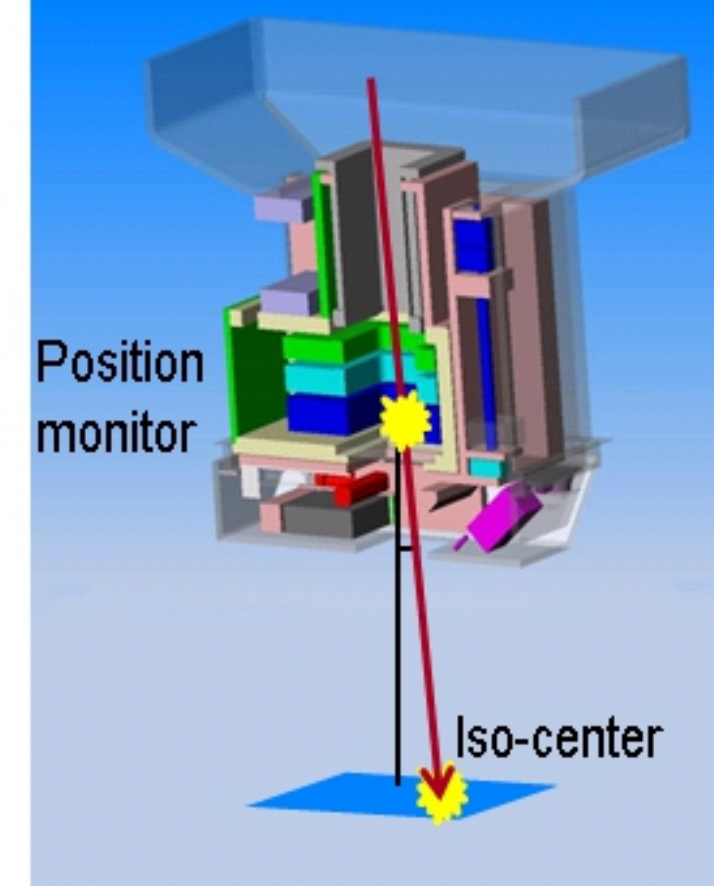}
\includegraphics[width=.51\linewidth]{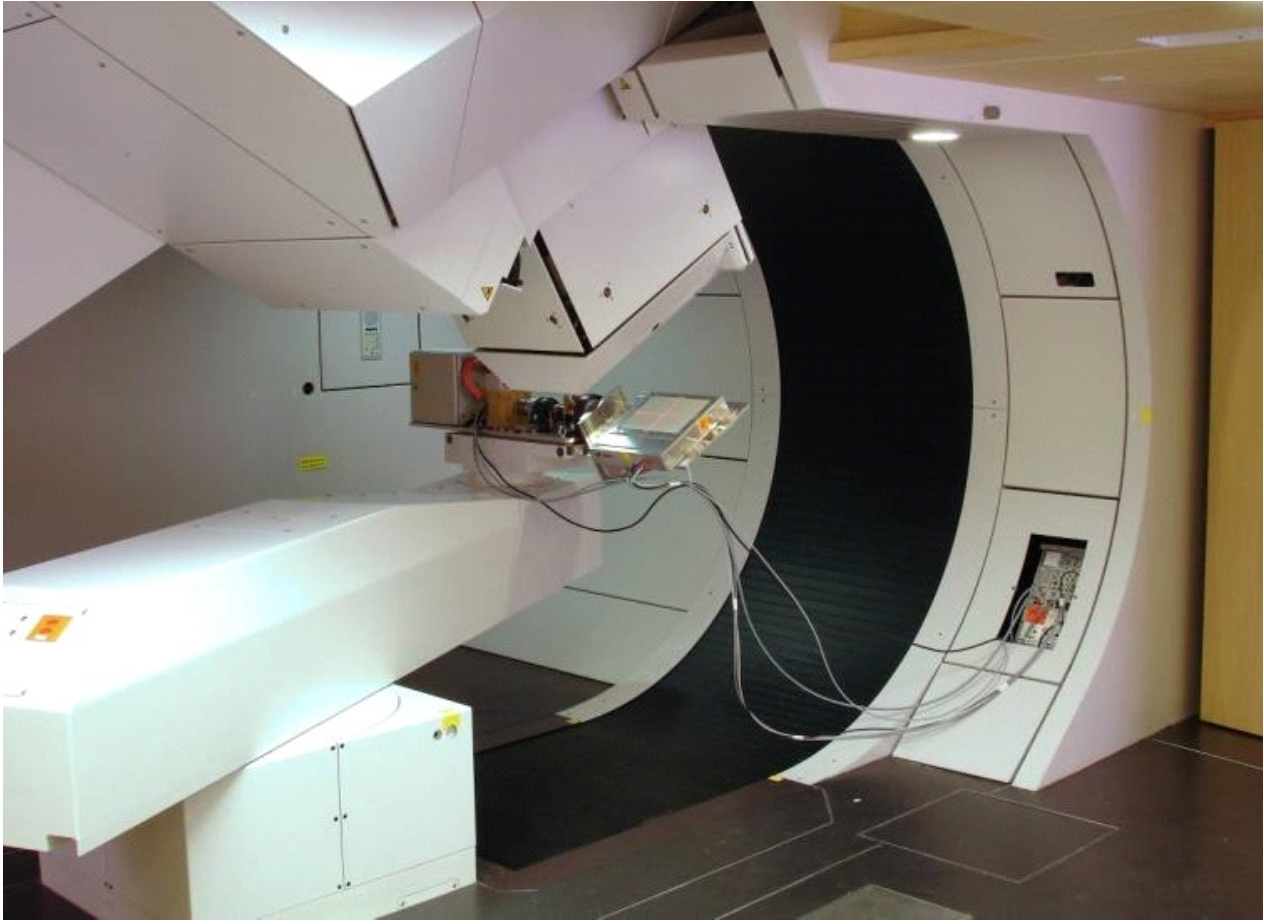}
\caption{Precise data on the beam angle is required to predict the position at the isocentre from the measured position in the nozzle chamber (left-hand side). At PSI, a second strip chamber was positioned at the isocentre to measure the correlation (right-hand side). The same setup is also used for the calibration of the scanner magnets.}
\label{fig:NozzleChamberCalibration.pdf}
\end{center}
\end{figure}

During dose delivery with active scanning, the position of the pencil beam is continuously monitored with a position monitor. Segmented ICs are the monitors of choice for minimizing the additional amount of material in the beam path. However, they measure the beam in the nozzle approximately \Unit{1}{m} before the isocentre and therefore the measured position needs to be projected to the isocentre. This requires profound knowledge of the beam angle for the full scanning range and all beam energies. An uncertainty in the beam angle of approximately \Unit{1}{mrad} already leads to a position measurement error of \Unit{1}{mm} (see \Fref{fig:NozzleChamberCalibration.pdf} (left-hand side)). The same measurement setup as for the scanner magnet calibration can be used.

\subsection{Spot size measurements}
\begin{figure}[ht]
\begin{center}
\includegraphics[width=.42\linewidth]{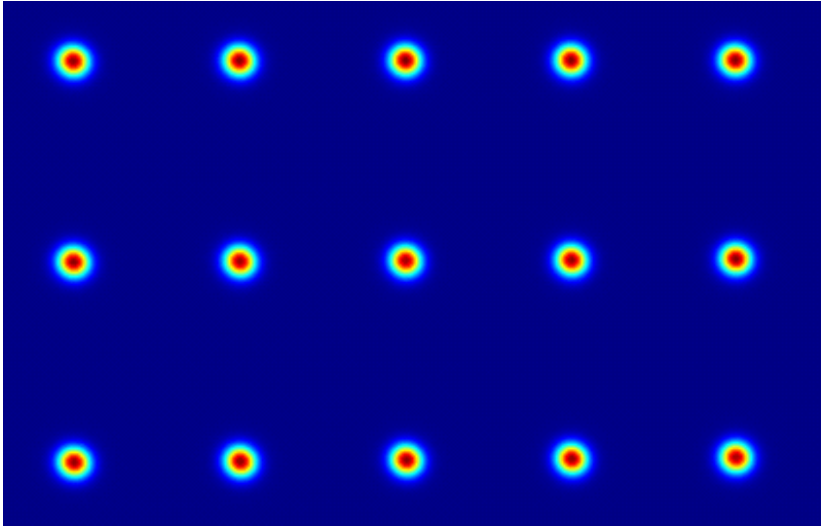}
\includegraphics[width=.4\linewidth]{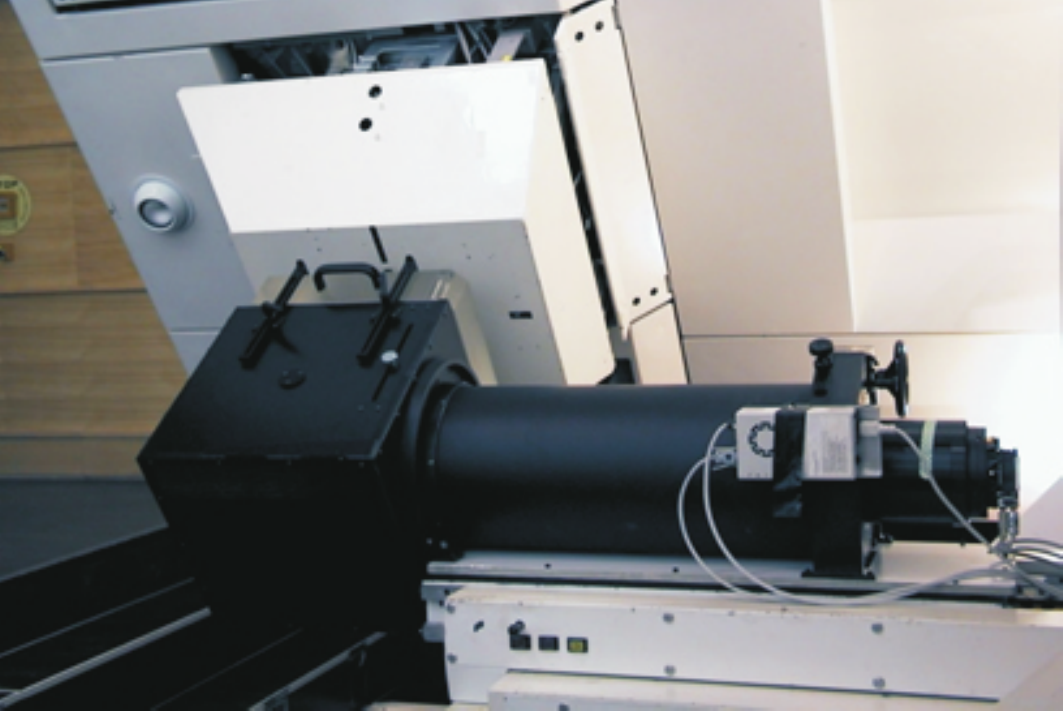}
\caption{Typical spot pattern for spot size determination for a beam energy of \Unit{150}{MeV} (left-hand side). The device for this measurement is based on a scintillating screen and a CCD camera (right-hand side).}
\label{fig:SpotSize.pdf}
\end{center}
\end{figure}

In addition to the depth--dose curve, the spot size is the other important input parameter for the TPS. The TPS can work with a single spot size for a given energy as long as spot size variations are negligible over the entire scan range. To assess this assumption a two-dimensional spot shape measurement needs to be performed at the isocentre over the full scan and proton energy range. At PSI, we are using a system which is based on a CCD camera looking at a scintillating foil, which provides high spatial resolution (see \Fref{fig:SpotSize.pdf} (right-hand side)). Similar commercial equipment is available, such as the Lynx detector from IBA Dosimetry in Schwarzenbruck (Germany), for example. The spot size can be extracted from a delivered spot pattern as shown in \Fref{fig:SpotSize.pdf} (left-hand side).
Multiple spots in one single data acquisition increase the measurement efficiency as long as the spots are well separated and the tails of the spots are not overlapping. A two-dimensional Gaussian is fitted for each spot to determine the spot size, and the required parameters for the TPS are extracted, depending on the beam model in the TPS.

\section{Dose monitor calibration}
For a correct dose distribution, the exact dose per spot determination is as essential as the spatial location of the spot, and the number of delivered protons must be detected with high accuracy. ICs with a minimal amount of material are again the preferred measurement devices because the dose monitor stays permanently in the beam path. The collected charge from the IC
\begin{equation}
Q_{\mathrm{IC}} = G_{\mathrm{IC}} \cdot S(E) \cdot \frac{1}{k_{Tp}} \cdot q_{e} \cdot n_{p}
\end{equation}
is proportional to the number of protons $n_{p}$ and the stopping power $S(E)$ for a given energy. The factor $k_{Tp}$ corrects for temperature and pressure other than standard conditions. The theoretical calculated value for the IC gain $G_{\mathrm{IC}}$ based on chamber geometry has large uncertainties and therefore experimental determination is required. At PSI, we use a Faraday cup for this calibration. The charge measured with the Faraday cup is $Q_{\mathrm{FC}}= q_{e} n_{p}$ and can be used for direct gain determination
\begin{equation}
G_{\mathrm{IC}} = \frac{Q_{\mathrm{IC}} \cdot k_{Tp}} {S(E) \cdot Q_{\mathrm{FC}}}~.
\end{equation}
There is also an alternative calibration procedure based on a measurement with a small IC~\cite{Goma2014}.

In general, dose distributions are calculated with pencil beam models which predict the dose per incident proton~\cite{Pedroni2004b}. The IC calibration is needed to convert the number of protons from a pencil beam to the expected IC signal. The validity of the IC calibration is verified by measuring the dose delivered to a water phantom with a certified thimble IC following a code of practice~\cite{Andreo2000}. At PSI, we found an agreement of 2--3\% between the dose calculated with the beam model and absolute dosimetry measurements. An additional empirical correction factor reduces this difference to less than 0.5\%.

\section{Summary and conclusion}
It is far beyond the scope of this article to give a complete description of medical commissioning. Important topics such as calibration of the imaging systems or the patient table were not touched. The goal of this text is to illustrate the methodology for some of the major tasks.

After successful medical commissioning, the system is validated by acceptance tests with similar characteristics and is then ready for patient treatment. After clinical operation has started, the performance of the system is continuously monitored by quality assurance (QA) tests. These tests have different repetition periods ranging from daily to yearly tests. The careful execution of a QA program is necessary to guarantee clinical operation with constant high quality over many years.

\section*{Acknowledgements}
The author would like to thank all members of the Center for Proton Therapy at PSI who contributed to this work.

\end{document}